\documentclass[showpacs,showkeys,preprintnumbers]{revtex4}%
\usepackage{amsfonts}
\usepackage{amsmath}
\usepackage{graphicx}
\usepackage{dcolumn}
\usepackage{bm}
\usepackage{amssymb}%
\begin{document}
\title{The Glueball in a Chiral Linear Sigma Model with Vector Mesons}
\author{Stanislaus Janowski$^{\text{(a)}}$, Denis Parganlija$^{\text{(a)}}$, Francesco
Giacosa$^{\text{(a)}}$, and Dirk H.\ Rischke$^{\text{(a,b)}}$}
\affiliation{$^{\text{(a)}}$Institute for Theoretical Physics, Goethe University,
Max-von-Laue-Str.\ 1, D--60438 Frankfurt am Main, Germany }
\affiliation{$^{\text{(b)}}$Frankfurt Institute for Advanced Studies, Goethe University,
Ruth-Moufang-Str.\ 1, D--60438 Frankfurt am Main, Germany }

\begin{abstract}
We present a two-flavour linear sigma model with global chiral symmetry and
(axial-)vector mesons as well as an additional glueball degree of freedom. We
study the structure of the well-established scalar resonances $f_{0}(1370)$
and $f_{0}(1500)$: by a fit to experimentally known decay widths we find that
$f_{0}(1370)$ is predominantly a $\bar{q}q$ state and $f_{0}(1500)$ is
predominantly a glueball state. The overall phenomenology of these two
resonances can be well described. Other assignments for our mixed
quarkonium-glueball states are also tested, but turn out to be in worse
agreement with the phenomenology. As a by-product of our analysis, the gluon
condensate is determined.

\end{abstract}

\maketitle

\section{Introduction}

Glueballs, the bound states of the gauge bosons of QCD, the gluons, are
naturally expected in QCD due to the nonabelian nature of the theory: the
gluons interact strongly with themselves and thus they can bind and form
colorless states, analogously to what occurs in the quark sector. The
existence of glueballs has been studied in the framework of the effective bag
model for QCD already four decades ago \cite{bag-glueball} and it has been
further investigated in a variety of approaches \cite{Close}. Numerical
calculations of the Yang-Mills sector of QCD also find a full glueball
spectrum in which the scalar glueball is the lightest state \cite{Morningstar}.

Glueballs can mix with quarkonium ($\bar{q}q$) states with the same quantum
numbers. This makes the experimental search for glueballs more complicated,
because physical resonances emerge as mixed states. The scalar sector
$J^{PC}=0^{++}$ has been investigated in many works in the past. The resonance
$f_{0}(1500)$ is relatively narrow when compared to other scalar-isoscalar
states: for this reason it has been considered as a convincing candidate for a
glueball state. Mixing scenarios in which two quark-antiquark isoscalar states
$\bar{n}n$ and $\bar{s}s$ and one scalar glueball $gg$ mix and generate the
physical resonances $f_{0}(1370),$ $f_{0}(1500)$, and $f_{0}(1710)$ have been
discussed in Refs.\ \cite{refs,longglueball}.

The aim of this work is to extend the linear chiral model of
Refs.\ \cite{Paper1,Susanna} by including the dilaton/glueball field. The
first attempt to incorporate a glueball into a linear sigma model was
performed long ago in Ref.\ \cite{schechter}. The novel features of the
present study are the following: (i) The glueball is introduced as a dilaton
field within a theoretical framework where not only scalar and pseudoscalar
mesons, but also vector and axial-vector mesons are present from the very
beginning. This fact allows also for a calculation of decays into vector
mesons. The model is explicitly evaluated for the case of $N_{f}=2,$ for which
only one scalar-isoscalar quarkonium state exists: $\sigma\equiv\bar{n}n$
which mixes with the glueball. The two emerging mixed states are assigned to
the resonances $f_{0}(1370)$ which is, in accordance with Ref.\ \cite{Paper1},
predominantly a $\bar{q}q$ state, and with $f_{0}(1500)$ which is
predominantly a glueball state. (ii) We consequently test --- to our knowledge
for the first time --- this mixing scenario above 1 GeV in the framework of a
chiral model.

The model under consideration is built in accordance with the symmetries of
the QCD Lagrangian. It possesses the known degrees of freedom of low-energy
QCD [(pseudo)scalar and (axial-)vector mesons] as well as the same global
chiral invariance. Another feature of the QCD Lagrangian is scale (or
dilatation) invariance $x^{\mu}\rightarrow\lambda^{-1}x^{\mu}$ (where $x^{\mu
}$ is a Minkowski-space coordinate and $\lambda$ the scale parameter of the
conformal group). It is realized at the classical level but broken at the
quantum level due to the loop corrections in the Yang-Mills sector (scale
anomaly). In this work the breaking of scale invariance is implemented at
tree-level by means of a dilaton field (representing a glueball) with the
usual logarithmic dilaton potential \cite{schechter}. However, all the other
interaction terms (with the exception of the chiral anomaly) are
dilatation-invariant in the chiral limit.

Having constructed the Lagrangian of the effective model, we calculate the
masses of the pure $\bar{q}q$ and glueball states in the $J^{PC}=0^{++}$
channel, study their mixing and calculate the decay widths of the mixed
states. Although we work with $N_{f}=2,$ the use of flavor symmetry enables us
to calculate the decay widths of the scalar resonances into kaons and into
both the $\eta$ and $\eta^{\prime}$ mesons which contain the $s$-quark in
their flavor wave functions. After the study of the already mentioned scenario
where $f_{0}(1370)$ and $f_{0}(1500)$ are predominantly quarkonium and
glueball, respectively, we also test the alternative scenario in which the
resonance $f_{0}(1710)$ is predominantly glueball and scenarios in which
$f_{0}(600)$ is predominantly quarkonium. They, however, lead to
inconsistencies when compared to the present data and are therefore regarded
as less favorable.

This paper is organized as follows. In Sec.\ II the Lagrangian of the linear
sigma model with (axial-)vector and glueball degrees of freedom is
constructed. In Sec.\ III we discuss the results for the masses of the
quarkonium-glueball mixed states and their decay widths. In Sec.\ IV we
present our conclusions.

\section{The Model}

The Yang-Mills (YM) sector of QCD (QCD without quarks) is classically
invariant under dilatations. This symmetry is, however, broken at the quantum
level. The divergence of the corresponding current is the trace of the
energy-momentum tensor $T_{\text{YM}}^{\mu\nu}$ of the YM Lagrangian
\begin{equation}
\left(  T_{\text{YM}}\right)  _{\mu}^{\mu}=\frac{\beta(g)}{4g}\, G_{\mu\nu
}^{a}G^{a,\mu\nu} \neq0\text{ ,} \label{ta}%
\end{equation}
where $G_{\mu\nu}^{a}$ is the field-strength tensor of the gluon fields,
$g=g(\mu)$ is the renormalized coupling constant at the scale $\mu$, and the
$\beta$-function is given by $\beta(g)=\partial g/\partial\ln\mu.$ At the
one-loop level $\beta(g)=-bg^{3}$ with $b=11N_{c}/(48\pi^{2}).$ This implies
$g^{2}(\mu)=\left[  2b\ln(\mu/\Lambda_{YM})\right]  ^{-1},$ where
$\Lambda_{YM}\simeq200$ MeV is the Yang-Mills scale. A finite energy scale
thus emerges in a theory which is classically invariant under dilatation
(dimensional transmutation). The expectation value of the trace anomaly does
not vanish and represents the so-called gluon condensate:
\begin{equation}
\left\langle T_{YM,\mu}^{\mu}\right\rangle =-\frac{11N_{c}}{48}\left\langle
\frac{\alpha_{s}}{2\pi}\,G_{\mu\nu}^{a}G^{a,\mu\nu} \right\rangle
=-\frac{11N_{c}}{48}C^{4}\text{ ,} \label{gc}%
\end{equation}
where
\begin{equation}
C^{4}\simeq(300-600\text{ MeV})^{4}\text{ .}%
\end{equation}
The numerical values have been obtained through QCD sum rules (lower range of
the interval) \cite{Sumrules} and lattice simulations (higher range of the
interval) \cite{Lattice}.

At the composite level one can build an effective theory of the YM sector of
QCD by introducing a scalar dilaton field $G$ which describes the trace
anomaly. The dilaton Lagrangian reads \cite{schechter}%
\begin{equation}
\mathcal{L}_{dil}=\frac{1}{2}(\partial_{\mu}G)^{2}-\frac{1}{4}\frac{m_{G}^{2}%
}{\Lambda^{2}}\left(  G^{4}\ln\left\vert \frac{G}{\Lambda}\right\vert
-\frac{G^{4}}{4}\right)  \text{ .} \label{ldil}%
\end{equation}
The minimum $G_{0}$ of the dilaton potential is realized for $G_{0}=\Lambda.$
Upon shifting $G\rightarrow G_{0}+G,$ a particle with mass $m_{G}$ emerges,
which is interpreted as the scalar glueball. The numerical value has been
evaluated in Lattice QCD and reads $m_{G}\sim1.5$ GeV \cite{Morningstar}. The
logarithmic term of the potential explicitly breaks the invariance under a
dilatation transformation. The divergence of the corresponding current reads
$\partial_{\mu}J^{\mu}_{dil}=T_{dil,\,\mu}^{\;\mu} =-\frac{1}{4}m_{G}%
^{2}\Lambda^{2}$. This can be compared with the analogous quantity in
Eq.\ (\ref{gc}) which implies $\Lambda=\sqrt{11}\,C^{2}/(2m_{G})$.

QCD with quarks is also classically invariant under dilatation transformations
in the limit of zero quark masses (chiral limit). The scale of all hadronic
phenomena is given by the previously introduced energy scale $\Lambda_{YM}.$
This fact holds true also when the small but nonzero values of the quark
masses is considered. In order to describe these properties in a hadronic
model we now extend the linear sigma model with $U(N_{f})_{R}\times
U(N_{f})_{L}$ of Refs.\ \cite{Paper1,GG,Mainz,UBW} by including the dilaton.
To this end, the following criteria are applied \cite{dynrec}: (i) With the
exception of the chiral anomaly, the parameter $\Lambda$ from Eq.\ (\ref{ldil}%
), which comes from the Yang-Mills sector of the theory in accordance with
QCD, is the only dimensionful parameter of the Lagrangian in the chiral limit.
(ii) The Lagrangian is required to be finite for every finite value of the
gluon condensate $G_{0}.$ This, in turn, also assures that no singular terms
arise in the limit $G_{0}\rightarrow0.$ In accordance with the requirements
(i) and (ii) only terms with dimension exactly equal to 4 are allowed in the
chiral limit.

The hadronic Lagrangian obeying these requirements reads
\begin{align}%
\mathcal{L}%
&  =\mathcal{L}_{dil}+\text{\textrm{Tr}}\left[  (D^{\mu}\Phi)^{\dag}(D_{\mu
}\Phi)-m_{0}^{2}\left(  \frac{G}{G_{0}}\right)  ^{2}\Phi^{\dag}\Phi
-\lambda_{2}(\Phi^{\dag}\Phi)^{2}\right]  -\lambda_{1}(\text{\textrm{Tr}%
}\left[  \Phi^{\dag}\Phi\right]  )^{2}\nonumber\\
&  +c[\text{\textrm{det}}(\Phi^{\dag})+\text{\textrm{det}}(\Phi
)]+\text{\textrm{Tr}}\left[  H\left(  \Phi^{\dag}+\Phi\right)  \right]
-\frac{1}{4}\text{\textrm{Tr}}\left[  (L^{\mu\nu})^{2}+(R^{\mu\nu})^{2}\right]
\nonumber\\
&  +\frac{m_{1}^{2}}{2}\left(  \frac{G}{G_{0}}\right)  ^{2}\text{\textrm{Tr}%
}\left[  (L^{\mu})^{2}+(R^{\mu})^{2}\right]  +\frac{h_{1}}{2}\text{\textrm{Tr}%
}[\Phi^{\dag}\Phi]\text{\textrm{Tr}}[L_{\mu}L^{\mu}+R_{\mu}R^{\mu}]\nonumber\\
&  +h_{2}\text{\textrm{Tr}}[\Phi^{\dag}L_{\mu}L^{\mu}\Phi+\Phi R_{\mu}R^{\mu
}\Phi^{\dag}]+2h_{3}\text{\textrm{Tr}}[\Phi R_{\mu}\Phi^{\dag}L^{\mu
}]+...\text{ ,} \label{Lagrangian}%
\end{align}
where $\Phi$ represents the $N_{f}\times N_{f}$ (pseudo)scalar multiplet, and
$L^{\mu}$ and $R^{\mu}$ the left- and right-handed vector multiplets. The dots
represent further terms which do not affect the processes studied in this work
and can be omitted.

In the particular case of $N_{f}=2$ studied in this manuscript one has
$\Phi=(\sigma+i\eta_{N})\,t^{0}+(\vec{a}_{0}+i\vec{\pi})\cdot\vec{t}$ (our eta
meson $\eta_{N}$ contains only non-strange degrees of freedom), $L^{\mu
}=(\omega^{\mu}+f_{1}^{\mu})\,t^{0}+(\vec{\rho}^{\mu}+\vec{a}_{1}^{\mu}%
)\cdot\vec{t}$ and $R^{\mu}=(\omega^{\mu}-f_{1}^{\mu})\,t^{0}+(\vec{\rho}%
^{\mu}-\vec{a}_{1}^{\mu})\cdot\vec{t}$ ; $t^{0}$, $\vec{t}$ are the generators
of $U(2)$. Moreover, $D^{\mu}\Phi=\partial^{\mu}\Phi-ig_{1}(L^{\mu}\Phi-\Phi
R^{\mu})$, $L^{\mu\nu}=\partial^{\mu}L^{\nu}-\partial^{\nu}L^{\mu}$,
$R^{\mu\nu}=\partial^{\mu}R^{\nu}-\partial^{\nu}R^{\mu}$.

The explicit breaking of the global chiral symmetry is described by the term
\textrm{Tr}$[H(\Phi+\Phi^{\dagger})]\equiv h\sigma$ $(h=const$. $\propto
m_{q}^{2})$, which allows us to take into account the nonvanishing value
$m_{q}$ of the quark mass. This term contains the dimensionful parameter $h$
with $[h]$ $=[\mathrm{energy}^{3}]$ and also explicitly breaks the dilatation
invariance, just as the quark masses do in the underlying QCD Lagrangian.
Finally, the chiral anomaly is described by the term $c\,(\det\Phi+\det
\Phi^{\dagger})$ \cite{Hooft}. For $N_{f}=2$ the parameter $c$ carries the
dimension $[\mathrm{energy}^{2}]$ and represents a further breaking of
dilatation invariance. This term, however, is driven by instantons, which are
also a property of the Yang-Mills sector of QCD.

The identification of the fields of the models with the resonances listed in
Ref.\ \cite{PDG} is straightforward in the pseudoscalar and (axial-)vector
sectors: the fields $\vec{\pi}$ and $\eta_{N}$ correspond to the pion and the
$SU(2)$ counterpart of the $\eta$ meson, respectively, $\eta_{N}%
\equiv(\overline{u}u+\overline{d}d)/\sqrt{2}$, with a mass of about $700$ MeV.
This value can be obtained by `unmixing' the physical $\eta$ and $\eta
^{\prime}$ mesons, which also contain $\overline{s}s$ contributions. The
fields $\omega^{\mu}$ and $\vec{\rho}^{\;\mu}$ represent the $\omega(782)$ and
$\rho(770)$ vector mesons, respectively, while the fields $f_{1}^{\mu}$ and
$\vec{a}_{1}^{\;\mu}$ represent the $f_{1}(1285)$ and $a_{1}(1260)$
axial-vector mesons, respectively. (In principle, the physical $\omega$ and
$f_{1}$ states also contain $\overline{s}s$ contributions but their admixture
is small.) As shown in Ref.\ \cite{Paper1}, the $\sigma$ field should be
interpreted as a predominantly $\bar{q}q$ state because its decay width
decreases as $1/N_{c}$ in the limit of large number of colors. The $\sigma$
and $G$ fields mix: the physical fields $\sigma^{\prime}$ and $G^{\prime}$ are
obtained through an $SO(2)$ rotation, as we shall show in the following. Then
the first and most natural assignment is $\{\sigma^{\prime},G^{\prime
}\}=\{f_{0}(1370),f_{0}(1500)\},$ see Sec.\ IIIA. Note that the $\vec{a}_{0}$
state is assigned to the physical $a_{0}(1450)$ resonance in accordance with
results of Ref.\ \cite{Paper1}. Other assignments for $\{\sigma^{\prime
},G^{\prime}\}$ will be also tested in Secs.\ III.B and III.C and turn out to
less favourable.

In order to study the non-vanishing vacuum expectation values (vev's) of the
two $J^{PC}=0^{++}$ scalar-isoscalar fields of the model $\sigma$ and $G$, we
set all the other fields in Eq.\ (\ref{Lagrangian}) to zero and obtain:%
\begin{equation}%
\mathcal{L}%
_{\sigma G}=\mathcal{L}_{dil}+\frac{1}{2}(\partial^{\mu}\sigma)^{2}-\frac
{1}{2}\left[  m_{0}^{2}\left(  \frac{G}{G_{0}}\right)  ^{2}-c\right]
\sigma^{2}-\frac{1}{4}\left(  \lambda_{1}+\frac{\lambda_{2}}{2}\right)
\sigma^{4}+h\sigma\text{ .} \label{Lagrangian1}%
\end{equation}
Upon shifting the fields by their vev's, $\sigma\rightarrow\sigma+ \phi$ and
$G \rightarrow G+ G_{0}$, we obtain the masses of the states $\sigma=(\bar
{u}u+\bar{d}d)/\sqrt{2}$ and $G=gg$ \cite{Paper1},
\begin{equation}
M_{\sigma}^{2}=m_{0}^{2}-c+3\left(  \lambda_{1}+\frac{\lambda_{2}}{2}\right)
\phi^{2}\;,\;\;\;\;\;M_{G}^{2}=m_{0}^{2}\, \frac{\phi^{2}}{G_{0}^{2}}%
+m_{G}^{2}\, \frac{G_{0}^{2}}{\Lambda^{2}}\, \left(  1+3\ln\left\vert
\frac{G_{0}}{\Lambda}\right\vert \right)  \text{ .} \label{M_G}%
\end{equation}
Note that the pure glueball mass $M_{G}$ depends also on the quark condensate
$\phi$, but correctly reduces to $m_{G}$ in the limit $m_{0}^{2}=0$
(decoupling of quarkonia and glueball). In the presence of quarkonia,
$m_{0}^{2} \neq0$, the vev $G_{0}$ is given by the equation
\begin{equation}
-\frac{m_{0}^{2}\phi^{2}\Lambda^{2}}{m_{G}^{2}}=G_{0}^{4}\ln\left\vert
\frac{G_{0}}{\Lambda}\right\vert \text{.} \label{G0}%
\end{equation}

The shift of the fields by their vev's introduces a bilinear mixing term
$\sim\sigma G$ in the Lagrangian (\ref{Lagrangian1}). The physical fields
$\sigma^{\prime}$ and $G^{\prime}$ can be obtained through an $SO(2)$
rotation,
\begin{equation}
\left(
\begin{array}
[c]{c}%
\sigma^{\prime}\\
G^{\prime}%
\end{array}
\right)  =\left(
\begin{array}
[c]{cc}%
\cos\theta & \sin\theta\\
-\sin\theta & \cos\theta
\end{array}
\right)  \left(
\begin{array}
[c]{c}%
\sigma\\
G
\end{array}
\right)  \text{ ,}%
\end{equation}
with%
\begin{align}
M_{\sigma^{\prime}}^{2}  &  =M_{\sigma}^{2}\cos^{2}\theta+M_{G}^{2}\sin
^{2}\theta+2\,m_{0}^{2}\,\frac{\phi}{G_{0}}\sin(2\theta)\text{ ,}%
\label{m_sigma_r}\\
M_{G^{\prime}}^{2}  &  =M_{G}^{2}\cos^{2}\theta+M_{\sigma}^{2}\sin^{2}%
\theta-2\, m_{0}^{2}\,\frac{\phi}{G_{0}}\sin(2\theta)\text{ ,} \label{m_G_r}%
\end{align}
where the mixing angle $\theta$ reads
\begin{equation}
\theta=\frac{1}{2}\arctan\left[  - 4\, \frac{\phi}{G_{0}}\,\frac{m_{0}^{2}}{
M_{G}^{2} -M_{\sigma}^{2} }\right]  \text{ .} \label{theta}%
\end{equation}
The quantity $m_{0}^{2}$ can be calculated from the masses of the pion,
$\eta_{N}$ and the bare $\sigma$ mass [see Ref.\ \cite{Paper1}]:
\begin{equation}
m_{0}^{2}=\left(  \frac{m_{\pi}}{Z}\right)  ^{2}+\frac{1}{2}\left[  \left(
\frac{m_{\eta_{N}}}{Z}\right)  ^{2}-M_{\sigma}^{2}\right]  \text{.}
\label{m_0}%
\end{equation}
If $m_{0}^{2}-c<0$, spontaneous breaking of chiral symmetry is realized.

\section{Results and Discussion}

The Lagrangian (\ref{Lagrangian}) contains the following twelve free
parameters: $m_{0}$, $\lambda_{1}$, $\lambda_{2}$, $m_{1}$, $g_{1}$, $c$, $h$,
$h_{1}$, $h_{2}$, $h_{3}$, $m_{G}$, $\Lambda=\sqrt{11}\,C^{2}/(2m_{G})$. The
processes that we will consider depend only on the combination $h_{1}%
+h_{2}+h_{3},$ thus reducing the number of parameters to ten. We replace the
set of ten parameters by the following equivalent set: $m_{\pi}$, $m_{\eta
_{N}}$, $m_{\rho}$, $m_{a_{1}}$, $\phi$, $Z$, $M_{\sigma}$, $m_{G}$, $m_{1}$,
$C$. The masses $m_{\pi}$ ($=139.57$ MeV) and $m_{\rho}$ ($=775.49$ MeV) are
fixed to their PDG values.

As outlined in Refs.\ \cite{Paper1, Mainz}, the mass of the $\eta_{N}$ meson
can be calculated using the mixing of strange and non-strange contributions in
the physical fields $\eta$ and $\eta^{\prime}(958)$:
\begin{equation}
\eta=\eta_{N}\cos\varphi+\eta_{S}\sin\varphi,\text{ }\eta^{\prime}=-\eta
_{N}\sin\varphi+\eta_{S}\cos\varphi\text{,} \label{phi}%
\end{equation}
where $\eta_{S}$ denotes a pure $\bar{s}s$ state and $\varphi\simeq-36^{\circ
}$ \cite{Giacosa:2007up}. In this way, we obtain the value $m_{\eta_{N}}=716$
MeV. (Given the well-known uncertainty of the value of the angle $\varphi$,
one could also consider other values, e.g., $\varphi=-41.4^{\circ}$, as
published by the KLOE Collaboration \cite{KLOE}, which corresponds to
$m_{\eta_{N}}=755$ MeV. Variations of the pseudoscalar mixing angle affect the
results presented in this paper only slightly.)

The value of $m_{a_{1}}$ is fixed to $1050$ MeV according to the study of
Ref.\ \cite{UBW}. (We note that taking the present PDG estimate of 1230 MeV
does not change our conclusions.) The chiral condensate is fixed as
$\phi=Zf_{\pi}$ and the renormalization constant $Z$ is determined by the
study of the process $a_{1}\rightarrow\pi\gamma$: $Z=1.67\pm0.2$ \cite{Paper1}.

\subsection{Assigning $\sigma^{\prime}$ and $G^{\prime}$ to $f_{0}(1370)$ and
$f_{0}(1500)$}

The $\sigma^{\prime}$ field denotes an isoscalar $J^{PC}=0^{++}$ state and its
assignment to a physical state is a long-debated problem of low-energy QCD
\cite{Close,refs,Morningstar,fariborz}. The two major candidates are the
resonances $f_{0}(600)$ and the $f_{0}(1370)$ \cite{PDG}. The study of
Ref.\ \cite{Paper1} has shown that $f_{0}(1370)$ is favoured to be a state
which is predominantly $\bar{q}q$. As stated above, the resonance
$f_{0}(1500)$ is a convincing glueball candidate. For these reasons we first
test the scenario in which $\{\sigma^{\prime},G^{\prime}\}=\{f_{0}%
(1370),f_{0}(1500)\},$ which turns out to be phenomenologically successful,
see below.

We are left with the following four free parameters: $C,M_{\sigma},m_{G}%
,m_{1}.$ They can be obtained by a fit to the five experimental quantities of
Table \ref{Table1}: the masses of the resonances $f_{0}(1500)$ ($M_{G^{\prime
}}\equiv M_{f_{0}(1500)}=1505$ MeV \cite{PDG}) and $f_{0}(1370)$ (for which we
use the mean value $M_{\sigma^{\prime}}^{ex}= (1350\pm150)$ MeV taking into
account the PDG mass range between 1200 MeV and 1500 MeV \cite{PDG}) and the
three well-known decay widths of the well-measured resonance $f_{0}(1500)$:
$f_{0}(1500)\rightarrow\pi\pi$, $f_{0}(1500)\rightarrow\eta\eta$ and
$f_{0}(1500)\rightarrow K\bar{K}$.%

\begin{table}[h] \centering
\begin{tabular}
[c]{|c|c|c|}\hline
Quantity & Our Value [MeV] & Experiment [MeV]\\\hline
$M_{\sigma^{\prime}}$ & $1191\pm26$ & $1200$-$1500$\\\hline
$M_{G^{\prime}}$ & $1505\pm6$ & $1505\pm6$\\\hline
$G^{\prime}\rightarrow\pi\pi$ & $38\pm5$ & $38.04\pm4.95$\\\hline
$G^{\prime}\rightarrow\eta\eta$ & $5.3\pm1.3$ & $5.56\pm1.34$\\\hline
$G^{\prime}\rightarrow K\bar{K}$ & $9.3\pm1.7$ & $9.37\pm1.69$\\\hline
\end{tabular}%
\caption{Fit in the scenario \{$\sigma
'$, $G'$\} = \{$f_0(1370)$, $f_0(1500)$\}. Note that the $f_0(1370)$ mass ranges between 1200 MeV and
1500 MeV \cite{PDG}
and therefore, as an estimate, we are using the value $m_{\sigma'}%
=(1350 \pm150)$ MeV in the fit.\label{Table1}}%
\end{table}%

Using the Lagrangian (\ref{Lagrangian}), these observables can be expressed as
functions of the parameters listed above. Note that, although our framework is
based on $N_{f}=2$, we can calculate the amplitudes for the decays into mesons
containing strange quarks by making use of the flavor symmetry $SU(N_{f}=3)$
\cite{longglueball}. It is then possible to calculate the following
$f_{0}(1500)$ decay widths into pseudoscalar mesons containing $s$-quarks:
$f_{0}(1500)\rightarrow K\bar{K}$, $f_{0}(1500)\rightarrow\eta\eta$ and
$f_{0}(1500)\rightarrow\eta\eta^{\prime}$.

The $\chi^{2}$ method yields $\chi^{2}/$d.o.f. $=0.29$ (thus very small),
$C=(699\pm40)$ MeV, $M_{\sigma}=(1275\pm30)$ MeV, $m_{G}=(1369\pm26)$ MeV and
$m_{1}=(809\pm18)$ MeV.

The consequences of this fit are the following:

(i) The quarkonium-glueball mixing angle reads $\theta=\left(  29.7\pm
3.6\right)
{{}^\circ}%
$. This, in turn, implies that the resonance $f_{0}(1500)$ consists to $76\%$
of a glueball and to the remaining $24\%$ of a quark-antiquark state. An
inverted situation holds for $f_{0}(1370)$.

(ii) Our fit allows us to determine the gluon condensate: $C=(699\pm40)$ MeV.
This result implies that the upper value in Eq.\ (\ref{gc}) is favoured by our
analysis. It is remarkable that insights into this basic quantity of QCD can
be obtained from the PDG data on mesons.

(iii) Further results for the $f_{0}(1500)$ meson are reported in the first
two entries of Table \ref{Table2}. The decay into $4\pi$ is calculated as a
product of an intermediate $\rho\rho$ decay. To this end the usual integration
over the $\rho$ spectral function is performed. Our result yields 30 MeV in
the $4\pi$ decay channel and is about half of the experimental value
$\Gamma_{f_{0}(1500)\rightarrow4\pi}$ $=54.0\pm7.1$ MeV. However, it should be
noted that an intermediate state consisting of two $f_{0}(600)$ mesons (which
is also expected to contribute in this decay channel) is not included in the
present model. The decay into the $\eta\eta^{\prime}$ channel is also
evaluated; this channel is subtle because it is exactly on the threshold of
the $f_{0}(1500)$ mass. Therefore, an integration over the spectral function
of the decaying meson $f_{0}(1500)$ is necessary. The result is in a
qualitative agreement with the experiment.

(iv) The results for the $f_{0}(1370)$ meson are reported in the last four
rows of Table \ref{Table2}. They are in agreement with the experimental data
regarding the full width: $\Gamma_{f_{0}(1370)}=(200$-$500)$\ MeV \cite{PDG}.
Unfortunately, the experimental results in the different channels are not yet
conclusive. Our theoretical results point towards a dominant direct $\pi\pi$
and a non-negligible $\eta\eta$ contribution; these results correspond well to
the experimental analysis of Ref.\ \cite{Bugg} where $\Gamma_{f_{0}%
(1370)\rightarrow\pi\pi}=325$ MeV and $\Gamma_{f_{0}(1370)\rightarrow\eta\eta
}/\Gamma_{f_{0}(1370)\rightarrow\pi\pi}=$ $0.19\pm0.07$ are obtained. We find
that the four-pion decay of $f_{0}(1370)\rightarrow\rho\rho\rightarrow4\pi$ is
strongly suppressed (as was also found in Ref.\ \cite{Paper1}). This is unlike
Ref.\ \cite{Bugg}, where a small but non-negligible value of about $50$ MeV is
found. However, it should be noted that due to interference effects our result
for this decay channel varies strongly when the parameters are even slightly modified.

(v) The mass of the $\rho$ meson can be expressed as $m_{\rho}^{2}=m_{1}%
^{2}+\phi^{2}\left(  h_{1}+h_{2}+h_{3}\right)  /2.$ In order that the
contribution of the chiral condensate is not negative, the condition
$m_{1}\leq m_{\rho}$ should hold. In the framework of our fit this condition
is fulfilled at the two-sigma level. This result points towards a dominant
$m_{1}$ contribution to the $\rho$ mass. This property, in turn, means that
the $\rho$ mass is predominantly generated from the gluon condensate and not
from the chiral condensate. It is therefore expected that the $\rho$ mass in
the medium scales as the gluon condensate rather than as the chiral
condensate. In view of the fact that $m_{1}$ is slightly larger than $m_{\rho
}$ we have also repeated the fit by fixing $m_{1}=m_{\rho}$: the minimum has a
$\chi^{2}/$d.o.f. $\simeq1$ and the results are very similar to the previous
case. The corresponding discussion about the phenomenology is unchanged.

(vi) As already stressed in Refs.\ \cite{Paper1,Zakopane}, the inclusion of
(axial-)vector mesons plays a central role to obtain the present results. The
artificial decoupling of (axial-)vector states would generate a by far too
wide $f_{0}(1370)$ state. For this reason the glueball-quarkonium mixing
scenario above 1 GeV has been previously studied only in phenomenological
models with flavor symmetry \cite{Close,longglueball} but not in the context
of chirally invariant models.%

\begin{table}[h] \centering
\begin{tabular}
[c]{|c|c|c|}\hline
Quantity & Our Value [MeV] & Experiment [MeV]\\\hline
$G^{\prime}\rightarrow\rho\rho\rightarrow4\pi$ & $30$ & $54.0\pm7.1$\\\hline
$G^{\prime}\rightarrow\eta\eta^{\prime}$ & $0.6$ & $2.1\pm1.0$\\\hline
$\sigma^{\prime}\rightarrow\pi\pi$ & $284\pm43$ & -\\\hline
$\sigma^{\prime}\rightarrow\eta\eta$ & $72\pm6$ & -\\\hline
$\sigma^{\prime}\rightarrow K\bar{K}$ & $4.6\pm2.1$ & -\\\hline
$\sigma^{\prime}\rightarrow\rho\rho\rightarrow4\pi$ & $0.09$ & -\\\hline
\end{tabular}%
\caption{Further results regarding the $\sigma' \equiv
f_0(1370)$ and $G' \equiv f_0(1500)$ decays.\label{Table2}}%
\end{table}%

\subsection{Assigning $\sigma^{\prime}$ and $G^{\prime}$ to $f_{0}(1370)$ and
$f_{0}(1710)$}

Although the resonance $f_{0}(1710)$ has also been regarded as a glueball
candidate in a variety of works \cite{refs2}, its enhanced decay into kaons
and its rather small decay width make it compatible with a dominant ${\bar{s}%
}s$ contribution in its wave function. Nonetheless, we have also tested the
assumption that the pure quarkonium and glueball states mix to produce the
resonances $f_{0}(1370)$ and $f_{0}(1710)$.

The resonance $f_{0}(1710)$ is experimentally well known. Decays into $\pi
\pi,$ $\bar{K}K$, and $\eta\eta$ have been seen, while no decay into $\eta
\eta^{\prime}$ and into $4\pi$ have been detected. Using the total decay width
$\Gamma_{f_{0}(1710)}= (135\pm8)$ MeV and the branching ratios reported in
Ref.\ \cite{PDG} it is possible to deduce the decay widths into $\pi\pi,$
$\bar{K}K$, and $\eta\eta,$ see Table III.

A fit analogous to the one in Table I yields too large errors for the decay
width $\sigma^{\prime}\equiv f_{0}(1370)\rightarrow\pi\pi.$ For this reason we
repeat our fit by adding the following constraint: $\Gamma_{\sigma^{\prime
}\rightarrow\pi\pi}=(250\pm150)$ MeV. The large error assures that this value
is in agreement with experimental data on this decay width. The results of the
fit are reported in Table III.%

\begin{table}[h] \centering
\begin{tabular}
[c]{|c|c|c|}\hline
Quantity & Our Value [MeV] & Experiment [MeV]\\\hline
$M_{\sigma^{\prime}}$ & $1386\pm134$ & $1350\pm150$\\\hline
$M_{G^{\prime}}$ & $1720\pm6$ & $1720\pm6$\\\hline
$G^{\prime}\rightarrow\pi\pi$ & $29.7\pm6.5$ & $29.3\pm6.5$\\\hline
$G^{\prime}\rightarrow\eta\eta$ & $6.9\pm5.8$ & $34.3\pm17.6$\\\hline
$G^{\prime}\rightarrow K\bar{K}$ & $16\pm14$ & $71.4\pm29.1$\\\hline
$\sigma^{\prime}\rightarrow\pi\pi$ & $379\pm147$ & $250\pm150$\\\hline
\end{tabular}%
\caption{Fit in the scenario \{$\sigma
'$, $G'$\} = \{$f_0(1370)$, $f_0(1710)$\}.
\label{SzenarioIII}}%
\end{table}%

\bigskip

The $\chi^{2}$ fit yields $\chi^{2}/$d.o.f. $=1.72$, $C=(764\pm256)$ MeV,
$M_{\sigma}=(1516\pm80)$ MeV, $m_{G}=(1531\pm233)$ MeV and $m_{1}=(827\pm36)$
MeV. The mixing angle between the pure quarkonium $\sigma$ and the pure
glueball $G$ calculated from Eq.\ (\ref{theta}) is $\theta=37.2%
{{}^\circ}%
\pm21.4%
{{}^\circ}%
$. The $\chi^{2}$ is wors than in the previous case, but the overall agreement
is acceptable. The mixing angle is large and could also overshoot the value of
$45%
{{}^\circ}%
,$ which would imply a somewhat unexpected and unnatural reversed ordering, in
which $f_{0}(1370)$ is predominantly glueball and $f_{0}(1710)$ predominantly quarkonium.

In Table IV we report the decay widths $G^{\prime}\rightarrow4\pi$,
$G^{\prime}\rightarrow\eta\eta^{\prime},$ $\sigma^{\prime}\rightarrow\eta\eta$
and $\sigma^{\prime}\rightarrow K\bar{K}$ , which can be calculated as a
consequence of the fit of Table III.%

\begin{table}[h] \centering
\begin{tabular}
[c]{|l|l|c|}\hline
Decay Width & Our Value [MeV] & Experimental value [MeV]\\\hline
$G^{\prime}\rightarrow4\pi$ & $115$ & -\\\hline
$G^{\prime}\rightarrow\eta\eta^{\prime}$ & $16$ & -\\\hline
$\sigma^{\prime}\rightarrow\eta\eta$ & $153\pm79$ & -\\\hline
$\sigma^{\prime}\rightarrow K\bar{K}$ & $2.1_{-2.1}^{+13.6}$ & -\\\hline
\end{tabular}%
\caption{Further results from the fit with \{$\sigma
'$, $G$'\} = \{$f_0(1370)$, $f_0(1710)$\}.\label{SzenarioIII1}}%
\end{table}%

\bigskip

A clear problem in the framework of this scenario emerges: the decay width
$G^{\prime}\equiv f_{0}(1710)\rightarrow4\pi$ is large, while experimentally
it has not been seen. Therefore, we conclude that this scenario is not
favoured. Moreover, in this scenario the remaining resonance $f_{0}(1500)$
should then be interpreted as a predominantly $\bar{s}s$ state, contrary to
what its experimentally dominant $\pi\pi$ decay pattern suggests.
Consequently, $f_{0}(1710)$ is unlikely to be predominantly a glueball state;
this is also in accordance with the results from the ZEUS Collaboration
\cite{ZEUS}.

\subsection{Scenarios with $\sigma^{\prime}\equiv f_{0}(600)$}

The scenarios $\{\sigma^{\prime},G^{\prime}\}=$ $\{f_{0}(600),$ $f_{0}%
(1500)\}$ and $\{\sigma^{\prime},G^{\prime}\}=$ $\{f_{0}(600),$ $f_{0}%
(1710)\}$ have also been tested. In both cases the mixing angle turns out to
be small ($\lesssim15^{\circ}$), thus the state $f_{0}(600)$ is predominantly
quarkonium. Then, in these cases the analysis of Ref.\ \cite{Paper1} applies:
a simultaneous description of the $\pi\pi$ scattering lengths and the
$\sigma\rightarrow\pi\pi$ decay width cannot be achieved. For these reasons
the mixing scenarios with the resonance $f_{0}(600)$ as a quarkonium state are
not favoured.

\section{Conclusions and Outlook}

We have presented a globally chirally invariant linear sigma model with
(axial-)vector mesons and a dilaton/glueball degree of freedom. We have
studied the phenomenology of the scalar states for the case $N_{f}=2$: in the
favoured scenario the resonance $f_{0}(1500)$ is predominantly a glueball with
a subdominant $\bar{q}q$ component and, conversely, $f_{0}(1370)$ is
predominantly a quark-antiquark $(\bar{u}u+\bar{d}d)/\sqrt{2}$ state with a
subdominant glueball contribution. It is interesting to observe that the
success of the phenomenological description of these scalar resonances is due
to the inclusion of the (axial-)vector mesons in the model. The gluon
condensate is also an outcome of our study and turns out to be in agreement
with lattice QCD results. Different scenarios in which $f_{0}(1710)$ is
predominantly glueball and/or $f_{0}(600)$ is predominantly quarkonium do not
seem to be in agreement with the present experimental data.

Natural extensions of the model are the case $N_{f}=3$ \cite{Krakau} and the
inclusion of a nonet of tetraquark states as additional low-lying scalar
states. In this general scenario, a mixing of five scalar-isoscalar states
takes place, which allows to describe all relevant scalar-isoscalar resonances
listed in the PDG below 1.8 GeV \cite{PDG}. Applications of the model at
nonzero temperature and density are also important because the presence of the
dilaton field allows to study the restoration of both the dilatation and the
chiral symmetry of QCD.

\section*{Acknowledgments}

The authors acknowledge fruitful discussions with A.\ Heinz and E.\ Seel. The
work of D.P.\ is supported by the Foundation "Polytechnical Society."

\end{document}